\documentclass{article}
\usepackage[utf8]{inputenc}
\usepackage{authblk}
\usepackage{hyperref}
\usepackage{bm}
\usepackage{xcolor}
\usepackage{graphicx}
\usepackage{setspace}
\usepackage{braket}
\usepackage{amsmath}
\usepackage{amssymb}
\usepackage{amsfonts}
\usepackage{physics}
\usepackage{mathtools}

\title {Navigating the Conjectural Labyrinth of the Black Hole Information Paradox}
\author{Galina Weinstein}
\affil{\normalsize Reichman University, The Efi Arazi School of Computer Science, Herzliya; University of Haifa, The Department of Philosophy, Haifa, Israel.} 

\begin{document}

\maketitle

\begin{abstract}
This paper explores the enduring black hole information and firewall paradoxes, challenges that have prompted many proposals, conjectures, and theories. Noteworthy among these are the ER $=$ EPR conjecture and AdS/CFT correspondence, which suggest possible avenues toward the yet-to-be-realized unified theory of quantum gravity. 
This discourse offers a comprehensive analysis of the aforementioned paradoxes, drawing upon insights from efforts to reconcile the schism between general relativity and quantum mechanics.    
\end{abstract}

\tableofcontents 

\section{Introduction}

General relativity has been accepted for over a century, but some anomalies challenge its validity. One significant anomaly comes from the discrepancy between quantum mechanics' principle of unitarity and the no-hair theorem and the equivalence principle of general relativity, particularly when looking at Hawking radiation from black holes. This conflict leads to the information and firewall paradox, discussed in sections \ref{1} and \ref{3}.

Resolving the information loss paradox requires finding a consistent framework that reconciles the unitarity of quantum mechanics with the no-hair theorem of general relativity. Various proposals and conjectures, such as the stretched horizon and complementarity, $A = R_B$ and the firewall, the ER $=$ EPR proposal (the connection between Einstein-Rosen bridges and entanglement or Einstein-Rosen-Podolsky pairs), AdS/CFT (a mathematical equivalence between a gravity theory in anti-de Sitter space-time and a conformal quantum field theory on its boundary), and complexity theory, have been explored to address this paradox and find a satisfactory resolution, as discussed in sections \ref{2}, \ref{3}, and \ref{4}.
However, these attempts have not yet led to a paradigm shift, which would involve the development of a comprehensive theory of quantum gravity capable of replacing both general relativity and quantum field theory. Such a theory would encompass and explain the limitations of both theories.

It is conceivable that AdS/CFT could play a role in this future theory, potentially forming a part of the new framework that replaces the current ones. Regarding reconciling general relativity and quantum mechanics, AdS/CFT provides a powerful tool known as holographic duality. It is holographic since the gravitational theory lives in (at least) one extra dimension \cite{Hartman}. According to the AdS/CFT correspondence, a theory of gravity in AdS space-time is mathematically equivalent to a quantum field theory (with no gravity) living on the boundary of that space-time. In other words, via the AdS/CFT correspondence, the CFT is also encodes the dynamics of the bulk (the interior of the AdS space), where gravity operates. 

Another perspective to consider AdS/CFT is looking at this duality through the following lens: AdS can be likened to "gravity in a box," where the boundary of the box (AdS space) accommodates a CFT, which subsequently encodes the gravitational dynamics occurring within the box \cite{Kaplan}. 

However, we must provide a clear and accurate qualification for these statements: we either do CFT or have an AdS spacetime, never both simultaneously. The theories are considered entirely
equivalent: any physical (gauge-invariant) quantity that can be computed in one theory can also be computed in the dual \cite{Hartman}. This duality allows researchers to study certain aspects of quantum gravity using well-understood quantum field theories.

While AdS/CFT has been successful in providing insights into the nature of quantum gravity and has been extensively studied in the context of AdS space-time, its direct application to our universe is more challenging because AdS/CFT primarily deals with AdS space-time. In contrast, our universe is believed to be closer to dS space-time. AdS space-time has a negative cosmological constant, while dS space-time has a positive one. The duality between gravity in dS space-time and a quantum field theory on its boundary is not fully understood. However, some researchers have proposed extensions or modifications of AdS/CFT that could potentially apply to dS space-time.

The black hole information paradox has puzzled physicists and computer scientists for years. Daniel Harlow and Patrick Hayden's work shows that combining computational complexity with quantum mechanics might offer a new perspective. They suggest that getting information out of a black hole before it vanishes might be almost impossible, even with quantum computers. While this doesn't solve everything, it gives us a fresh perspective on the problem. As I show in section \ref{6}, today, many in the scientific community, especially computer scientists, see the blend of computation and quantum mechanics as a promising path forward for understanding black holes and their intricate information dynamics.

As I discuss in Section \ref{5}, Albert Einstein had already acknowledged the inconsistency between his general theory of relativity and quantum mechanics. He saw the need for a unified field theory grounded in space-time and gravity rather than on a quantum basis. Einstein did not focus on black holes, believing anything beyond the Schwarzschild singularity lacked physical significance. From his viewpoint, the anomaly lay in what he saw as problems in quantum mechanics, specifically entanglement, which is now widely accepted as a fundamental aspect of quantum mechanics. The major conflicts between his general relativity and quantum mechanics emerged after his death, especially concerning what he disliked: the Schwarzschild singularity.

As I show in section \ref{5}, the consensus shifted towards unifying gravity within the quantum framework after advances in quantum mechanics, a change evident in modern theories like the AdS/CFT correspondence. Recent theories revolve around quantum mechanics as foundational, linking to quantum information and computation, as demonstrated by the unitary transformation principle. The debates around the black hole information paradox have been pivotal, with theories like the fuzzball proposition, remnants, and the final burst offering explanations, each with unique implications and criticisms, and further discussions like the Page Time Paradox hinting at unresolved complexities in our understanding of black hole physics.

\section{The black hole information paradox and its spinoffs} \label{1}

\subsection{Hawking's black hole information paradox}

Let us commence our discussion with Stephen Hawking's definition: "The no hair theorem implied that all information about the collapsing body was lost from the outside region apart from three conserved quantities: the mass, the angular momentum, and the electric charge" \cite{Hawking2}.

Hawking initially suggested in 1975 that the process of black hole evaporation results in a loss of information about the state of particles that fell into the black hole, making the final state of the black hole and all the particles and radiation associated with it unpredictable if the black hole evaporated completely. If the black hole were to evaporate completely, the system's final state would be the state of all the radiation emitted, along with any remaining particles outside the black hole.

Hawking’s original calculations suggested that the emitted radiation would be in a mixed state due to the apparent loss of information about the particles that fell into the black hole. Unlike a pure state, described by a single vector in the Hilbert space, where we have complete knowledge about the system, a mixed state represents a lack of complete knowledge. 
This implied that a non-unitary S-matrix (scattering matrix) would describe the system's evolution (from infalling particles to outgoing radiation), signifying a departure from standard quantum mechanics where a unitary S-matrix describes processes. 
Unitarity ensures the total probability of all possible outcomes of any event always sums to 1, corresponding to the conservation of information. A unitary S-matrix ensures that probabilities are conserved in quantum processes. If the S-matrix is non-unitary, the evolution of the quantum system (the black hole and all the particles and radiation associated with it) is not deterministic, leading to unpredictability in the system's final state after the complete evaporation of the black hole. \cite{Hawking}.

\subsection{Hawking reformulates the information paradox}

Jacob Bekenstein and Hawking explored the idea that black holes can be associated with entropy, a measure of a system's disorder. Bekenstein calculated the amount of information a black hole could contain and expressed it in terms of its area (not the volume), which led to the formulation of black hole entropy. A black hole's information content can be quantified and linked to its entropy. 

In other words, if we describe mathematically how information is stored in qubits, the information of all the material that forms the black hole gets encoded (represented) on the black hole's event horizon (the boundary). 

According to Gerard 't Hooft, all phenomena within a volume can be described by degrees of freedom residing on the surface bounding that volume. This implies the world is a two-dimensional lattice of bits, with the bounding surface acting as a viewing screen onto which three-dimensional events are projected. 
Susskind proposed the holographic principle based on 't Hooft's ideas, positing that all the information contained within a volume of space can be encoded on a boundary surface surrounding that volume, much like a two-dimensional hologram can represent a three-dimensional object \cite{Susskind6}. 

The Bekenstein-Hawking formulation suggests this idea by demonstrating that a black hole's entropy (and hence information) scales with its surface area rather than its volume.

The holographic principle was formulated before discovering the AdS/CFT correspondence. The AdS/CFT correspondence later provided a concrete realization of the holographic principle within the context of string theory by establishing a precise duality between a gravity theory in a higher-dimensional space (AdS space) and a lower-dimensional quantum field theory (CFT) defined on its boundary. 

However, our observable universe appears more accurately described by dS space, complicating attempts to formulate a holographic description. In dS space, the boundary is at temporal infinity, making the definition of a holographic dual more elusive. The formulation and successful realization of holography in the context of AdS space via the AdS/CFT correspondence was a significant leap, as it showed how the holographic principle could be made precise and calculable. 

Within the AdS/CFT correspondence, a black hole in an AdS space can be described as a CFT living on the boundary of that space. This correspondence suggests that the evolution of black holes can be described by the evolution of a certain quantum system (described by the CFT) that doesn't have gravity. In quantum field theories, the evolution of systems is unitary, meaning that information is conserved over time. Through this lens, the process of black hole evaporation and the subsequent dynamics could be described in a unitary fashion, suggesting that all the information about the state of particles (matter) thrown into the black hole isn't lost but is encoded in the quantum state of a CFT living on the boundary of the AdS space in which the black hole resides. 

While Hawking initially proposed that information gets lost in black hole processes, the AdS/CFT correspondence suggests otherwise. Even though the details about the particles that fell into the black hole seem lost due to the black hole evaporation, this original form of the information is not truly lost, but instead encoded and transferred into the Hawking radiation. Extracting the information from the radiation would involve a process of decoding. The decoding process would hypothetically allow one to retrieve information about the initial state of the particles that fell into the black hole by analyzing the characteristics of the outgoing Hawking radiation.
	
In 2005, Hawking noted a shift in understanding regarding the black hole information paradox. He conceded that the AdS/CFT correspondence has shown no information loss. 
In other words, the evolution and evaporation of black holes can be described in a unitary, information-conserving manner through the dual quantum field theory description. This represents a reversal from Hawking's earlier stance, which suggested that information could be lost in black holes, thereby contributing to the black hole information paradox\cite{Hawking}. 
	
The shift in understanding, primarily brought about by the AdS/CFT correspondence, moved the discourse from information loss to the conservation of information and the notions of encoding and decoding the information in the context of black holes. However, despite this shift, resolving the information paradox remains difficult. The paradox now hinges on understanding how the information is encoded in the Hawking radiation and how, in theory, it could be decoded or retrieved.  However, the practicality and exact mechanism of the decoding are among the complex issues intertwined with the black hole information paradox.

\subsection{The firewall paradox}

An observer falling into a black hole would be in free fall. 
From the perspective of the falling observer, there would be no dramatic indication at the event horizon that they have crossed any special boundary. This is the "no drama" scenario.  
The transition would appear smooth, and they would continue freely falling inward.
The observer would not feel any sudden forces and disruptions when crossing the event horizon, assuming the black hole is sufficiently massive to have a weak tidal gradient at the horizon. The equivalence principle states that locally (in small enough regions of spacetime), the effects of gravity are indistinguishable from acceleration in a non-gravitational field. So, as the falling observer approaches and crosses the event horizon of a large black hole, the equivalence principle would still be valid locally. The observer falling into the black hole would feel like they were freely floating; even if they were to drop a small object next to them, it would fall at the same rate. 

When considering Hawking radiation over the lifespan of a black hole, early and late radiation would be entangled due to the continuous process of particle pair creation near the event horizon. Suppose the Hawking radiation is in a pure state. In that case, the total system, including the early and late radiation, must maintain an entanglement reflecting the quantum correlations established at their emission. Thus, the purity of Hawking radiation suggests that late radiation is fully entangled with early radiation. 
The lack of a dramatic event for an infalling observer indicates that the late radiation is fully entangled with modes behind the black hole horizon.

This scenario hints at a type of cloning that contradicts quantum mechanics principles, specifically the strong subadditivity of entropy. Strong subadditivity is a fundamental inequality related to quantum entropies which states that for any three quantum systems $A$, $B$, and $C$, the following inequality holds: $S_{AB}+S_{BC}\geq S_B+S_{ABC}$. 

The \emph{AMPS paradox}, or \emph{Firewall Paradox}, named after the authors Ahmed Almheiri, Donald Marolf, Joseph Polchinski, and James Sully, arises from analyzing the behavior of an old black hole (a black hole in the late stages of its evaporation). A contradiction arises if one assumes that the late Hawking radiation is entangled with the early Hawking radiation (to preserve quantum purity) and that the infalling observer encounters no drama. 
According to quantum mechanics, a quantum system (like a qubit) can be entangled with only one other system at a time. However, in this scenario, it would imply that the late Hawking radiation is entangled with both the early radiation and the modes inside the horizon, hinting at a sort of cloning of entanglement, which is not allowed. 

AMPS demonstrate an entropy violation. The entropy is decreasing for an old black hole, which violates the strong subadditivity condition and other derived relations. If entropy is decreasing, and given the scenario of the AMPS paradox described above, the strong subadditivity of entropy is violated. This violation hints at a deeper inconsistency in understanding the nature of black holes, Hawking radiation, and quantum entanglement \cite{AMPS}.

\section{The stretched horizon and complementarity} \label{2}

\subsection{The stretched horizon and the membrane paradigm}

In 1986, Douglas A. Macdonald, Richard H. Price, and Kip S. Thorne introduced the notion of a "stretched horizon" and the classical \emph{membrane paradigm}. The stretched horizon is a surface that is modeled to curve slightly outside the event horizon of a black hole \cite{Thorne}, \cite{Thorne2} (see discussion in \cite{Wallace1}, \cite{Wallace2}). 

According to the classical membrane paradigm, black holes are conceptualized as a two-dimensional surface or membrane. When describing a black hole, talking about a surface or a membrane is useful, even though black holes are not compact objects with a surface. The stretched horizon is a conceptual and mathematical construct rather than a real physical surface. By imagining a surface with membrane-like properties slightly outside the event horizon of a black hole, physicists can simplify certain complex problems and handle them more manageable. 

The stretched horizon idea facilitates the analysis of black hole dynamics and interactions, making the complex physics of black holes more tractable and comprehensible \cite{Wilczek}:

1) In Schwarzschild coordinates, the metric becomes singular (i.e., undefined) at the event horizon, leading to a coordinate singularity. Alternative coordinates like Kruskal-Szekeres and Eddington-Finkelstein coordinates are smooth through the event horizon and do not have this problem.
While these coordinate systems can eliminate the coordinate singularity at the event horizon, considering a surface slightly outside the true horizon, one can avoid the coordinate singularities. 

2) The event horizon is also the location of an infinite redshift: light emitted from just outside the horizon will be infinitely redshifted, i.e., its energy will be reduced to zero as seen by an observer at infinity. 

3) As one approaches the event horizon, tidal gravitational forces become infinitely large. However, for a sufficiently massive black hole, these forces can be small at the event horizon and diverge as one approaches the singularity at $r=0$. The curvature of space and the tidal gravitational forces are less severe near a massive black hole's event horizon than a small black hole. An observer falling into a massive black hole might not experience significant tidal forces until they reach the singularity at $r = 0$. 

4) Within the context of Schwarzschild coordinates, at the event horizon, the Jacobian matrix becomes singular (it loses its invertibility), and the transformation between different coordinate systems degenerates at that point. If the Jacobian is singular at the event horizon, it leads to a degenerate metric in the transformed coordinate system. When one switches to a coordinate system better suited for analyzing the geometry near the event horizon, such as Kruskal-Szekeres coordinates, the coordinate singularity at the event horizon is eliminated, and the metric tensor remains non-degenerate. The Jacobians do not become singular at the event horizon in these coordinates, and a well-behaved metric tensor can be obtained. 

The stretched horizon is chosen to be timelike. It has a well-defined geometry and doesn’t suffer from the singularities associated with a null event horizon. On a timelike surface, it is possible to have a family of time-like observers who hover just outside the true horizon. These observers are called "fiducial observers" and have a well-defined time experience. Unlike an observer at the true event horizon who would experience infinite gravitational time dilation, these fiducial observers experience a large but finite redshift relative to observers far from the black hole. 
Because it is timelike, the stretched horizon allows for a clear definition of boundary conditions and a non-degenerate, well-behaved, induced metric. 

In short, the stretched horizon is not only about overcoming coordinate singularities. It behaves more like ordinary space, so standard techniques from classical physics can be applied. 

\subsection{The four postulates of complementarity}

In 1993, Leonard Susskind, Lárus Thorlacius, and John Uglum introduced the concept of black hole complementarity \cite{Susskind1}. The stretched horizon becomes invisible to an observer who falls through it, demonstrating a complementarity between observations made by infalling and distant observers. By considering a 2D model of gravity, Susskind et al. outline the behavior, kinematics, and statistical fluctuations of the stretched horizon, emphasizing its importance in understanding black hole formation and evaporation while considering quantum mechanical principles.

Susskind et al. initially develop four postulates to establish the foundation of black hole complementarity \cite{Susskind1}, \cite{AMPSS}:

\emph{Postulate 1}. \emph{Unitarity}: The formation and evaporation of a black hole, as observed by a distant observer, can be entirely described within the framework of standard quantum theory. A unitary S-matrix describes the evolution from the infalling matter to the outgoing Hawking-like radiation. 

\emph{Postulate 2}. \emph{Effective field theory}: Outside the stretched horizon of a massive black hole, physics can be described as a good approximation by a set of semi-classical field equations. Although a consistent formulation of four-dimensional gravity is lacking, a simplified two-dimensional gravity model effectively illustrates a stretched horizon. 
These semi-classical equations, part field-theoretic and part thermodynamic describe the average energy flow and the evolution of the horizon. 

\emph{Postulate 3}: To a distant observer, a black hole appears to be a quantum system with discrete energy levels. The dimension of the subspace of states describing a black hole of mass $M$ is the exponential of the Bekenstein entropy $S_{BH}(M)$ without any infinite additive constant in the entropy. 

While Susskind et al. outline the above three postulates focusing on the perspective of an outside observer, they also bring up an additional assumption concerning the experience of an infalling observer as per the equivalence principle. This assumption is a form of a fourth postulate, although not explicitly labeled as such in the text. It contrasts with Postulate 1, as it hints at a divergence in the observed phenomena between an external observer and an infalling observer, particularly around the event horizon of a black hole: 

\emph{Postulate 4}. \emph{No drama}: A freely falling observer experiences nothing unusual when crossing the event horizon. The low-energy dynamics probed by this observer near their worldline are well-described by the Lorentz-invariant effective field theory (this theory ensures that the effective descriptions are consistent with the fundamental symmetries of the underlying quantum field theory). 

\subsection{The semi-classical stretched horizon}

Susskind and colleagues describe their "semiclassical stretched horizon" as being just one Planck unit larger in area than the global event horizon itself. They suggest that for an external observer to not reference events inside a black hole, a theory based on Postulates 1-3 is needed. To address this, they introduce the concept of a stretched horizon, a visible timelike curve positioned just ahead of the black hole's event horizon.
Susskind et al. state that Postulates 1-3 are best implemented by viewing the stretched horizon through a thermodynamic lens, which is assumed to have a microphysical foundation. Despite the coarse-grained thermodynamic description (which overlooks the minute details and focuses on broader, averaged properties), Susskind et. al. believe there is an underlying microphysical basis, i.e., a finer-grained level of description of the stretched horizon based on microscopic physical properties. This finer description would be more detailed and might explain the thermodynamic behaviors observed at the coarse-grained level regarding underlying quantum mechanical processes \cite{Susskind1}. 

In the finer-grained level of description of the stretched horizon, when information hits the stretched horizon, it gets scrambled and mixed up with other information, but it is not lost. The scrambling is so effective that it appears to an outside observer that the information has been destroyed, although, in principle, it has been conserved. Black holes evaporate over time, and as they do so, they re-emit the scrambled information. This re-emission of Hawking radiation is considered a quantum mechanical process, where pairs of particles are created near the horizon, one falling in and the other escaping, carrying away a bit of the black hole's mass and potentially some scrambled information. The stretched horizon paradigm addresses the dynamics of black holes at the quantum level by providing a framework to analyze the quantum gravitational effects near the black hole horizon. 

The properties and behaviors of the semi-classical stretched horizon are similar to a physical membrane with mechanical, electrical, and thermal characteristics, as seen by an outside observer. Due to the gravitational dynamics around a black hole, the stretched horizon is portrayed as extremely complex and chaotic, with features akin to a viscous fluid and thermal properties governed by its area. When something approaches the black hole, the membrane absorbs and scrambles the information before it crosses the event horizon and re-emits it. The idea that information is scrambled and re-emitted before reaching the event horizon contradicts the no-drama scenario. The stretched horizon surface could show dramatic interactions contradicting general relativity's no-drama scenario. The membrane paradigm posits that physics outside the black hole should not be affected by dynamics inside it. The formulation helps separate the regions inside and outside the black hole to focus on observable phenomena.\footnote{I refer to David Wallace's discussion to gain further insights into the membrane paradigm and its implications on black hole information loss \cite{Wallace}, \cite{Wallace1}, \cite{Wallace2}.}

\subsection{Black hole complementarity}

In 1993, Susskind et al. demonstrated that black hole complementarity reconciles two seemingly contradictory observations: the infalling observer and the outside observer. They posited that the experience of an observer falling into a black hole (no drama) and that of an observer remaining outside (who perceives the horizon as a dynamical membrane or stretched horizon and sees information scrambled and re-emitted) are both valid. However, these experiences cannot be reconciled due to the fundamental limitations imposed by the event horizon. In other words, the nature of the event horizon creates a fundamental division between the experiences of inside and outside observers.

Niels Bohr introduced the idea of complementarity to reconcile the wave-particle duality of quantum mechanics. Susskind et al. invoke the concept of complementarity in the context of understanding the nature of the stretched horizon and the quantum behaviors surrounding black holes. 

Susskind et al. suggest that different experimental setups reveal different aspects of quantum systems, and these aspects, though seemingly contradictory, are complementary and collectively contribute to the full understanding of the black hole information paradox. Like the wave-particle duality, they suggest that depending on the experiment conducted, one may detect a quantum membrane (stretched horizon) around a black hole. 
Both experiments provide valid but complementary information about the black hole's nature. They hint at a scenario where different experiments, akin to the different setups used to observe particle or wave behavior, will reveal either the quantum membrane aspect or some other aspect of black holes: "An experiment of one kind will detect a quantum membrane, while
an experiment of another kind will not." However, no observer can ascertain the results of both types of experiments simultaneously. The membrane is not an invariant that all observers can agree upon, indicating a level of complementary understanding required to reconcile different observational outcomes \cite{Susskind1}. \cite{Susskind2}. 

Susskind et al. also touch on the interaction of infalling matter with the atoms of the stretched horizon, leading to a thermal state and subsequent evaporation, hinting at the quantum mechanical interactions at play. The observed states and behaviors can be viewed as complementary descriptions of the black hole's behavior from different observational and theoretical frameworks \cite{Susskind1}, \cite{Polchinski}.

\section{The Firewall and complementarity} \label{3}

\subsection{Contradictions among the four postulates}

In 2013, AMPS evaluated the consistency of Susskind et al.'s four postulates regarding black hole complementarity in an "old" black hole scenario. They considered a black hole formed from a pure state and decaying over time; they analyzed its Hawking radiation and divided it into an early part and a late part, especially focusing on a scenario after the black hole has emitted half of its initial Bekenstein-Hawking entropy. AMPS show that their derivation contradicts Susskind et al.'s Postulates 1, 2, and 4 for an old black hole. 
According to Postulate 1, the state of Hawking radiation is pure. According to Postulate 2, the physics outside the stretched horizon can be described semi-classically, implying a predictable and "free" evolution of modes outside the horizon. By applying Postulates 1 and 2, AMPS conclude that an infalling observer would encounter high-energy modes. This contradicts Postulate 4, which posits that an infalling observer should experience nothing of the ordinary while crossing the horizon \cite{AMPS}. 

Let us explore the intricacies of AMPS' argument. Postulate 2 means low-energy dynamics near an observer's worldline still obey the familiar Lorentz-invariant field theory. Also, according to Postulate 4, an observer falling into a black hole is unlikely to come across extremely high-energy quanta. 
When considering black holes that form and decay, Postulate 1 indicates that the Hawking radiation from the black hole remains in a pure state. AMPS divide the Hawking radiation into an early part and a late part. After the Page time, when a black hole emitted half of its initial Bekenstein-Hawking entropy, the number of states in the early radiation vastly outnumbers the states in the late radiation. 
Postulate 1 implies that the state of Hawking radiation remains pure, even as it is divided between early and late radiation. 

AMPS write equations highlighting the interaction between outgoing Hawking modes of radiation and infalling modes, which lead to the creation of Hawking radiation. They show how these modes, especially the ones from late radiation, are experienced by an infalling observer. This is central to the firewall paradox because this derivation suggests that the late radiation is deeply entangled with the early radiation for an old black hole (one after the Page time). 
Because of this deep entanglement between early and late radiations, an observer falling into the black hole would encounter high-energy modes (a "firewall") as they approach and cross the event horizon. 
This contradicts Postulate 4, which asserts that an infalling observer would experience nothing unusual upon crossing the event horizon of the black hole.

AMPS use Eddington-Finkelstein coordinates to visually portray how the postulates, when applied to an old black hole scenario, lead to a contradiction concerning the experience of an observer falling into a black hole and the encounter with high-energy modes contrary to the postulates of black hole complementarity. 

The specific mode (mode $b$) that is part of the Hawking radiation emitted by the black hole starts as a well-defined Hawking photon at a large distance from the black hole (as per postulate 1). Still, as the observer encounters it nearer to the black hole, its wavelength is much shorter (much smaller than the Schwarzschild radius). In other words, the mode's frequency is much higher (or its wavelength much shorter) than one would expect near the event horizon of the black hole, contradicting Postulate 4. 

AMPS define $a^\dagger a$, representing a quantum measurement of the infalling modes, which are modes that an observer falling into the black hole would naturally measure. Suppose the observer's measurements show that the field is in an eigenstate of this number operator. In that case, it implies they detect a definite number of particles in the infalling mode, and it would contradict Postulate 1, which insists that the state of Hawking radiation is pure. $b^\dagger b$ is the number operator for the outgoing modes associated with the Hawking radiation emitted from the black hole. Suppose the observer's measurements align with an eigenstate of this operator. In that case, it contradicts Postulate 4, which states that a freely falling observer should experience nothing of the ordinary when crossing the horizon. If the results the observer obtains are contingent on when they fall into the black hole, then it would violate Postulate 2. This postulate posits that the results should be universal, regardless of when the observer falls in \cite{AMPS}. 

\subsection{A drama: the firewall as a physical barrier}

The main conclusion of AMPS’ paper is that if one assumes that the process of black hole evaporation is unitary and follows the postulates of black hole complementarity, a contradiction arises, leading to the unexpected result of a "firewall," a barrier to information retrieval from black holes, at the event horizon. This firewall would be a highly energetic boundary that would incinerate anything that attempts to cross it, contradicting the idea of a smooth horizon expected from classical general relativity.
The firewall would, therefore, very much be a "drama" for an infalling observer, Alice. According to AMPS, she would encounter the firewall as she approached the event horizon and would be immediately destroyed. On the other hand, the processes near the event horizon look very different for an outside observer, Bob. Bob, who stays safely outside the black hole and never falls in, will never see Alice cross the event horizon. Due to the extreme gravitational time dilation effects, as Alice gets closer and closer to the horizon, she appears to move more slowly. From Bob's viewpoint, Alice will asymptotically approach the horizon but never quite cross it. Alice would appear thermalized to Bob in the stretched-horizon hot Planckian layer just above the event horizon. She seems absorbed and radiated away as part of the black hole's Hawking radiation. From Bob's perspective, Alice appears to be destroyed by the firewall as she asymptotically approaches the horizon.
The firewall, therefore, behaves much like the stretched horizon in terms of its position relative to the black hole. However, while the stretched horizon is a benign region that an observer can pass through without noticing anything unusual, the firewall is a highly energetic and destructive barrier that an observer cannot pass through unscathed.

In their argument, AMPS examine the behavior of black holes past the Page time and the scrambling time. The scrambling time gives an insight into how quickly information can become hidden in the black hole's microstates.
When information enters a black hole, it gets mixed up with the chaotic state of information within it. The scrambling time is when this newly introduced information gets dispersed and mixed up to become virtually irretrievable from the previous chaos within the black hole. The quicker the scrambling time, the quicker the information gets dispersed within the black hole, making it harder or impossible to retrieve later. Scrambling time is described concerning black holes being the fastest scramblers in nature. Fast scramblers are systems that achieve a minimal bound on the scrambling time, indicating that they scramble information exceptionally quickly. Black holes are examples of fast scramblers due to their purported ability to scramble information at a rate that approaches the theoretically lower limit. This scrambling makes it exceedingly difficult to retrieve any information about the state of particles that have fallen into the black hole. 

AMPS conclude by stating that following the scrambling time, "we have to expect either drama for the infalling observer or novel physics outside the black hole \cite{AMPS}.

Recall that all black holes can be completely described by just three externally observable classical parameters: mass, electric charge, and angular momentum. Any other information about the matter that formed a black hole and that fell into it ("hair") is not accessible to outside observers. The no-hair theorem suggests that black holes reach a simple, stable configuration once they have settled down or aged. How long does it take for a black hole to lose its "hair," i.e., for information about in-falling matter to become scrambled and indiscernible across its horizon? One can liken this process to a localized perturbation that is introduced at a specific location on the black hole's stretched horizon, thereby temporarily disrupting the uniformity of the horizon. Over time, this localized disturbance diffuses and spreads out over the entire stretched horizon, returning the system to a state of thermal equilibrium. The time it takes for the perturbation to spread out and cover the entire stretched horizon uniformly is the scrambling time. This is the time it takes for a black hole to "become bald", i.e., lose any distinguishing features or "hair" and reach a state where it can only be described by its mass, charge, and angular momentum. But there isn't a mathematical proof for this identification \cite{Sekino}. 

The firewall would form around when the black hole emitted about half of its initial entropy in Hawking radiation, corresponding to the Page time. The logic behind this timing is based on the idea that after the Page time, the outgoing radiation begins to carry away information about the black hole's interior, leading to a potential violation of the no-cloning theorem of quantum mechanics if there's no dramatic change at the horizon like a firewall.

\subsection{A critique of the firewall}

In 2017, William Unruh and Robert Wald criticized the firewall that suggests violating the principle of equivalence, as it would entail a dramatic and locally detectable change for infalling observers at the horizon. The firewall proposition doesn't provide a clear mechanism for why or how a high-energy barrier would suddenly emerge at the horizon of an old black hole. Such a mechanism must be consistent with known physics, and any \emph{ad hoc} addition raises concerns about its physical legitimacy. 

Unruh and Wald suggest that the firewall proposal might be an overreaction to the information paradox. They assert that while the paradox is indeed puzzling, it doesn't necessarily mandate such a drastic revision of our understanding of black hole horizons, especially one that conflicts with well-established principles. They believe there are other potential resolutions to the information paradox that do not require the introduction of firewalls. They argue for the validity of quantum field theory in regions away from the Planck scale. They suggest that the semiclassical analysis of black hole evaporation doesn't truly violate quantum theory. 

Unruh and Wald find such suggestions neither plausible nor palatable. They critique those adjusting and modifying quantum mechanics to get around the information loss problem. They find it ironic that some researchers, in their quest to preserve the principles of quantum mechanics, propose modifications or radical alternatives that might undermine quantum mechanics. Such drastic measures could violate quantum theory in situations where it should reliably work \cite{Unruh}.

\section{Bits, qubits, and error-correcting codes} \label{6}

\subsection{Hayden and Preskill's mirror} 

Consider again black holes that have passed their Page time. Recall that these would be black holes that have radiated away more than half their initial entropy and are on their way to evaporating eventually. As black holes near the end of their Hawking radiation process, they would be considered old. In 2007, Patrick Hayden and John Preskill suggested that such old black holes behave more like an \emph{information mirror} rather than a black abyss that swallows and destroys information irretrievably. Hayden and Preskill outlined a scenario in which Alice throws her diary into the black hole; instead of being lost, the information is reflected in Hawking radiation. This is akin to how a mirror reflects light \cite{Hayden}.

Hayden and Preskill imagined Alice, who decided to throw a diary containing sensitive information into an old black hole to ensure its secrecy. They speculated that Bob, an expert with advanced recovery skills, might later be able to retrieve the information contained in the diary from the Hawking radiation emitted by the black hole as it slowly evaporates over a long period. 

They first described a deterministic model where the black hole’s internal dynamics operate predictably. In this model, when Alice's diary enters the black hole, it is transformed into a series of bits combined with the black hole's internal bit string. This combined string is then subjected to a known permutation, which is deterministic and known to Bob. With his advanced knowledge of black hole dynamics and the known permutation, Bob observes the bits emitted from the black hole via Hawking radiation. Once Bob receives a small number of bits more than the original size of the diary, he can decode the entire diary. The extra bits Bob needs to receive provide enough information, alongside his understanding of the permutation and the black hole's internal state, to deduce the diary's content. Initially, the black hole was considered a perfect medium to conceal and destroy the information in Alice’s diary. However, Hayden and Preskill suggest that with enough understanding of the black hole's internal dynamics and collecting a slightly larger amount of data, Alice's diary can be decoded and recovered, challenging the idea of secure destruction of information \cite{Hayden}.

Suppose Alice has quantum information encoded in a $k$-qubit quantum memory, $M$, with the quantum state represented in a Hilbert space of dimension $2^k$. Hayden and Preskill then introduce a third party, Charlie, with a reference system $N$, initially maximally entangled with Alice's memory $M$. Alice, aiming to destroy the information in $M$, throws it into an old black hole. The black hole's complex and destructive dynamics would render the information in $M$ irretrievable. Bob tries to extract a subsystem from the Hawking radiation emitted by the black hole, post-Alice's disposal act. 

Suppose the subsystem Bob extracts is maximally entangled with Charlie's reference system $N$. In that case, this entanglement implies that the quantum information initially in $M$ has somehow made its way out of the black hole via Hawking radiation. In other words, the above entanglement is a channel that allows the transfer of quantum information from $M$ out of the black hole. By utilizing this entanglement, Bob can recover the quantum information initially in $M$.

Suppose the initial state of $M$ was in a pure state (i.e., not entangled with any other reference system), and the subsystem Bob extracts is maximally entangled with $N$. In that case, it again suggests that the information from $M$ has made its way out of the black hole. The maximal entanglement with $N$ allows Bob to recover the exact state in $M$ in his chosen subsystem and retrieve the original quantum information Alice tried to destroy. 
In other words, in principle, in both cases, the result is that Bob can recover the quantum information Alice attempted to destroy by extracting a subsystem from the Hawking radiation, which is maximally entangled with $N$. Bob, wanting to recover Alice’s memory $M$, collects the Hawking radiation emitted by the black hole. Hayden and Preskill propose that by collecting the Hawking radiation and waiting for just a little more than $k$ qubits to be emitted, Bob would then have enough information to begin decoding the state of Alice's qubits. 

Information of infalling qubits (i.e., Alice's memory $M$) gets scrambled with the black hole's state and is not permanently lost inside black holes. Hence, information that falls into a black hole is later emitted via Hawking radiation and can, in principle, be recovered, thereby aligning with the quantum mechanical principle of information conservation. This goes against the traditional belief stemming from the no-hair theorem that, from Bob’s point of view, information is permanently lost in black holes \cite{Hayden}. 

By suggesting a mechanism for information retrieval in 2007, Hayden and Preskill provided a potential resolution to the black hole information loss paradox.  

Once Alice’s information from $M$ enters the black hole, it undergoes scrambling, wherein the information it carries becomes thoroughly mixed with the state of the black hole in a highly chaotic but deterministic, unitary evolution. This scrambling is hypothesized to happen extremely fast, making the information ostensibly lost, though it's conserved in a quantum mechanical sense. If the black hole is a perfect scrambler, then in principle, the information about Alice's memory $M$ is not destroyed but is distributed non-locally within the emitted Hawking radiation. 

In the Hayden-Preskill model, two different scenarios are considered to explore the dynamics of information retrieval from a black hole \cite{Hayden}:

1. In the first scenario discussed above, Alice encodes quantum information in a $k$-qubit quantum memory $M$ and throws it into the black hole while she stays outside. The argument proposes that if the black hole has evaporated past its halfway point, then Bob can retrieve the information from the emitted Hawking radiation shortly after Alice’s quantum memory enters the black hole.

2. Alice carries the quantum memory $M$ in the second scenario as she falls into the old black hole. Here, the principle of black hole complementarity is invoked to reconcile the difference in experiences between Alice and Bob. According to black hole complementarity, Alice and Bob will have fundamentally different, non-overlapping experiences. From Alice's perspective, she falls into the black hole with $M$. From Bob's perspective, he may retrieve the information from the Hawking radiation outside the black hole.

Let us now explore the variation of the first scenario, where Alice herself falls into the black hole carrying the quantum information $M$. 

Consider Alice once again. This time, she falls into the black hole with $M$ while Bob stays outside and tries to decode $M$ from the Hawking radiation emitted by the black hole. After the Page time, the black hole could swiftly return the quantum information it receives via Hawking radiation to Bob. This timely return means that from Bob's perspective, the information is retrieved and not lost to the black hole, thus not violating the principle of unitarity. While Alice sees the information fall into the black hole, Bob retrieves the information from the Hawking radiation, and neither can verify the other's experience. Recall that black hole complementarity posits no single global description of events but that the descriptions from Alice and Bob's perspectives are valid. 

According to Hayden and Preskill, black hole complementarity suggests that the information is not cloned. The idea of cloning in their scenario refers to the possibility of having two copies of the same quantum information $M$: one inside the black hole (with Alice) and one outside (recovered by Bob from the Hawking radiation). If the information could be recovered from the outside while still being on the inside, it would seemingly violate the no-cloning theorem. 

Hayden and Preskill show that once Alice falls into the black hole, she cannot communicate with Bob to compare her information with what Bob recovers from the Hawking radiation. Similarly, suppose Bob decides to jump into the black hole after recovering the information from the Hawking radiation. In that case, he won't be able to compare notes with Alice or communicate his findings to anyone outside the black hole. The conditions inside the black hole prevent such verification. This unverifiability is crucial to avoiding a direct violation of the no-cloning theorem \cite{Hayden}.

Hayden and Preskill's thought experiment proposed that information thrown into a black hole can be recovered quickly from the Hawking radiation, given that the black hole has already emitted more than half of its initial entropy. From a quantum mechanical point of view, the catch here is the potential violation of the no-cloning theorem. If Bob can reconstruct the state outside the black hole while Alice still possesses it inside it, it seems the state has been cloned. Black hole complementarity suggests that no single observer can see both the inside and outside of a black hole, and thus, for any observer, the information is either inside or outside, but never both. This prevents the apparent cloning in the Hayden-Preskill scenario. However, AMPS challenged black hole complementarity by pointing out a contradiction. Suppose one accepts the postulates used in their argument. In that case, it leads to the conclusion that there must be a firewall, a highly energetic barrier just outside the event horizon, which would burn anything falling into the black hole. This firewall would be at odds with the no-drama scenario, which predicts that free-falling observers should experience no special effects when crossing the event horizon of a large black hole. So, if one accepts the conclusions of the AMPS paper and still wants to maintain the consistency of quantum mechanics, then a firewall would seem to be required in the Hayden-Preskill scenario. 

\subsection{The \texorpdfstring{$A=R_B$}{A=R\_B} conjecture and the Hayden-Harlow protocol}

AMPS then posited that a sufficiently old black hole could have a firewall at its event horizon instead of having a smooth spacetime interior. The firewall forms at the Page time (when a black hole has emitted about half its particles and half the information). Post the Page time, the modes (or quantum states) just behind the black hole horizon (labeled as $A$) can only be entangled with modes in front of the horizon (labeled as $B$) if they are identified with the radiation that has been emitted (labeled as $R$). This is expressed as $A = R_B$ \cite{AMPS}, \cite{Susskind}. 

AMPS argue that qubit $B$, obtained from late radiation from the black hole, can't be entangled with both qubit $A$ (behind the horizon) and qubit $R$ (from early radiation) due to the monogamy of entanglement, leading to the conclusion that there must be a firewall at the horizon where Alice would encounter high-energy particles instead of smoothly crossing the event horizon. 
The idea of the firewall contradicts the traditional understanding that a freely falling observer should experience nothing unusual as she crosses the horizon (due to the equivalence principle). 

Daniel Harlow and Hayden showed that the procedure Alice would have to undertake to verify the entanglement between qubits $B$ and $R$ or $R_B$ (and thereby concluding about the entanglement between qubits $B$ and $A$) would be computationally infeasible within the time frame before she falls into the black hole.

Harlow and Hayden use a computational complexity argument to suggest that decoding the information contained in the Hawking radiation is computationally infeasible for Alice until the black hole has almost completely evaporated. This argument posits a form of a decoding task that would be essential for Alice to perform to extract the information about Charlie's state from the Hawking radiation. Harlow and Hayden's decoding process is set within the AdS/CFT framework, exploring the practicality and computational feasibility of extracting information from Hawking radiation as envisioned by the AdS/CFT correspondence \cite{Harlow}. 

Harlow and Hayden’s approach attempts to address the practical feasibility of an infalling observer (Alice) being able to measure the Hawking radiation to distinguish between these entanglements before the black hole completely evaporates. They model an old black hole as a system of qubits and employ Page's theorem to represent the state of the black hole as old, which would be maximally mixed on a subsystem. Alice performs various quantum measurements before and after crossing the event horizon of a black hole. 

Consider the quantum state $\ket{\psi}$ of the black hole and its Hawking radiation. Utilizing Schmidt decomposition, the quantum state of the black hole and its radiation is represented with specific bases for two different subsystems, $B$ and $H$. The subsystems $B$, $H$, and $R$ represent different parts of $\ket{\psi}$. Subsystem $B$ is related to a specific mode in the near-horizon area of the black hole, $H$ represents the black hole horizon, and $R$ represents the radiation emitted from the black hole. A unitary transformation ($U_R$) relates the chosen bases (chosen for subsystems $B$ and $H$ within the quantum state $\ket{\psi}$) to the natural bases sets of orthogonal vectors that arise naturally when the quantum state is decomposed using the Schmidt procedure, demonstrating the entanglement between the subsystems $B$, $H$, and $R$. $U_R$ re-expresses $\ket{\psi}$ in terms of the natural bases arising from the Schmidt decomposition. $U_R$ transitions the state representation from the chosen bases to the natural bases, where the entanglement structure of $\ket{\psi}$ becomes more apparent. 

Alice needs to perform the transformation $U^{\dagger}_R$ using a quantum computer to distill the purification of subsystem $B$ from the radiation before she can observe a violation of entanglement monogamy as per the AMPS experiment.\footnote{Distillation in the context of quantum entanglement is akin to purification. Both processes aim to obtain a set of purer or more highly entangled states from a set of less entangled or mixed states. The analogy with distillation or purification in chemistry is often made to help intuitively understand the process. Just like in a chemical process where you might distill a liquid to remove impurities and obtain a purer substance, in quantum information theory, you would distill (or purify) a mixed quantum state to obtain a state of higher entanglement, which is often more useful for various quantum information processing tasks. The terms "distillation" and "purification" convey the idea of extracting and isolating the most useful part of something from a less pure or mixed state.} 

This distillation process aims to obtain a clearer and more usable form of subsystem $B$ from the radiation, which is essential for further analysis. Once the purification is distilled, Alice can enter the black hole to make the necessary observations. However, due to the extreme computational complexity of implementing the unitary transformation $U^{\dagger}_R$, doing so would take longer than the lifetime of the black hole itself. While the black hole can produce certain states without requiring exponential time, distilling $B$ from the radiation poses a significant computational challenge. Alice won’t have enough time to complete the necessary computations and observations to detect a violation of entanglement monogamy before she crosses the event horizon, which would result in her inability to obtain the desired information or validate/invalidate the AMPS paradox (a firewall). 

A circuit ($U_{dyn}$) can produce $\ket{\psi}$ in polynomial time. Even with an efficient $U_{dyn}$, the inverse of this circuit cannot be used to decode Hawking radiation due to the challenges in acting on degrees of freedom ($H$) inside the black hole. This highlights a boundary in our understanding and manipulation of black hole information. This inability is a prominent issue in black hole physics because it implies a lack of a complete quantum mechanical description of its interior. The distilling transformation $U^{\dagger}_R$ is established, which could act solely on radiation. However, this is nonconstructive because it only proves there is such a transformation without providing a method to achieve it. This leaves a gap in understanding the complexity of the distillation process, which is crucial for further analyses. The absence of a constructive method to achieve the transformation $U^{\dagger}_R$ leaves a crucial aspect of the black hole quantum mechanics — the complexity of the distillation process — unexplored. The complexity of this process is pivotal for understanding the practicality and feasibility of distilling and analyzing quantum information from black hole radiation \cite{Harlow}, \cite{Harlow2} (see explanation in \cite{Weinstein}). 

\subsection{Quantum error-correcting codes and the holographic principle}

In quantum computing, quantum error correction is a set of techniques to maintain the integrity of quantum information against errors due to decoherence and other quantum noise. 
Drawing inspiration from these techniques, computer scientists and physicists have explored the idea that the AdS/CFT correspondence might be understood through the lens of quantum error correction. The essence is to map the bulk to the boundary like a quantum error-correcting code. The Ryu-Takayanagi (RT) formula and quantum error correction are a crucial bridge between these two areas. This formula links the entanglement entropy of a sub-region on the boundary with the entropy of the bulk, specifically tying it to the minimal surface in the bulk.  

This perspective hints at a deep connection between quantum computation, quantum error correction, and holography that behaves much like a quantum error-correcting code. Specifically, the low-energy Hilbert space $\mathcal{H}_{bulk}$ of bulk effective field theory has a boundary described by a Hilbert space, $\mathcal{H}_{boundary}$. When discussing the low-energy Hilbert space $\mathcal{H}_{bulk}$ of the bulk effective field theory, we refer to the AdS/CFT correspondence used to explore aspects of effective field theories. In other words, at low energies, the bulk behaves in a way that a boundary theory can effectively capture without gravity. The emphasis on low energy means that this correspondence breaks down at extremely high energies or distances of order the Planck length \cite{Kaplan}.

Chris Akers, Netta Engelhardt, Daniel Harlow, Geo Penington, and Shreya Vardhan believe that the exterior of the black hole, i.e., Hawking radiation, seems to carry information about the interior, challenging the classical notion of black holes as information sinks. The Hawking radiation that Bob, the outside observer, observes (which carries information) seems to conflict with Alice's uninterrupted fall into the black hole. Drawing on the AdS/CFT correspondence, which posits that the information about the inside of a black hole can be encoded on its boundary, Akers, Engelhardt, Harlow, Penington, and Vardhan proposed that the black hole interior can be represented as a kind of quantum error-correcting code on the boundary. 

A correspondence called the dictionary $V$ maps states in the low-energy Hilbert space $\mathcal{H}_{bulk}$ of bulk effective field theory into the boundary Hilbert space $\mathcal{H}_{boundary}$. The quantum error-correcting code is a process (isometric map) represented by $V$, which takes logical states in the logical Hilbert
space $\mathcal{H}_{bulk}$. It encodes them in the physical Hilbert space $\mathcal{H}_{boundary}$ (an isometry is a linear map which obeys $V^\dagger V = I$). This encoding is redundant, i.e., some of the quantum information about the bulk can be lost and inaccessible without affecting our ability to reconstruct the bulk's physics. $V$ has certain holographic properties that determine how information can be reconstructed.

Alice's fall into the black hole represents information being thrown into the bulk (inside the black hole). Bob, staying outside, only has access to the boundary (the event horizon and surroundings). From Alice's perspective, she continues to experience the interior when she falls into a black hole. But from Bob's perspective, this information is scrambled and appears as part of the Hawking radiation outside the black hole. Over time, more of Alice's reality seems to be externalized. 

Through this framework, Bob might recover and decode some information about Alice and the black hole's interior without ever entering it. This information would be encoded in the Hawking radiation he observes. However, the complexity remains exponential, suggesting that the decoding would be computationally infeasible and take an incredibly long time. While the information isn't truly lost, extracting it would be exceptionally complex. The computational requirements for decoding this scrambled information are so enormous that it would seem virtually impossible.

Hawking initially advocated black hole information loss; see section \ref{1}. His prediction was based on an effective description (i.e., effective field theory or semiclassical calculations). Crucially, Hawking radiation appeared to carry no information about the state of the matter that had fallen into the black hole. 
Black holes emitted Hawking radiation (which appeared to be thermal and thus information-free). Eventually, they evaporated, seemingly erasing any information about the matter that collapsed to form them in the first place.

The problem with Hawking's calculation was the apparent non-unitarity. On the other hand, although the CFT side of AdS/CFT involves effective field theory, it is a fully unitary description. Thus, if information gets lost in the bulk (like in a black hole), it should still be recoverable on the boundary. The challenge is understanding how.

The effective description appeared to show no information, and it was coarse-graining. The semiclassical calculation is an effective and coarse-grained description, i.e., it doesn't consider all possible fine details or quantum effects. The fact is that information was not lost but was in such a scrambled form that Hawking’s semiclassical calculations couldn't capture it. Hence, while the information might be recoverable in principle (due to the error-correcting properties of holography), doing so would be computationally prohibitive because of the immense complexity involved \cite{Akers}.

\section{Avoiding a firewall} \label{4}

\subsection{Typical and atypical black holes}

According to Juan Maldacena and Susskind, an AdS eternal black hole is equivalent to two entangled black holes connected by an ER bridge \cite{Susskind5}.

Eternal black holes in AdS space are typically considered to be in equilibrium. But in 2013, Donald Marolf and Joseph Polchinski discussed black holes formed by collapse in AdS/CFT, writing: “While our black holes have been explicitly formed by collapse, to high accuracy they resemble the eternal black holes that dominate any corresponding microcanonical ensemble. 

In other words, under certain conditions, the black holes formed by collapse resemble the eternal black holes to a high degree of accuracy in a certain microcanonical ensemble comprising all possible configurations of a black hole at a specified energy. Marolf and Polchinski clarify that the term "collapse" here is used in a broader sense than usual. Typically, collapse might refer to a rapid process where a massive celestial object collapses to form a black hole in a short time relative to the AdS time scale associated with AdS space. However, in this context, the term also encompasses much slower processes of collapse that may occur over exponentially long times. Marolf and Polchinski broaden the collapse concept to include slower processes. In the usual rapid collapse, the resultant black holes occupy a subspace of small entropy, which limits the resemblance to eternal black holes. By considering a broader range of collapse processes, including much slower ones, the resultant black holes exhibit a closer resemblance to eternal black holes. Thus, more varieties of black hole states may be included within the microcanonical ensemble of all possible configurations a black hole could have at a particular energy level. This would provide a richer set of black hole states for theoretical examination and analysis \cite{Polchinski2}.

However, the above analogy is limited. While it may provide a useful mathematical and theoretical framework for certain analyses, it doesn't fully capture the complexities and specifics of \emph{conventional} collapsing black hole formation and behavior. 
Conventional black holes formed by collapse, i.e., Schwarzschild and Kerr black holes, are considered "typical." The typical nature of the black hole is understood in terms of how these black holes are formed through known processes (e.g., stellar collapse) in our universe. 
On the other hand, AdS eternal black holes are theoretical constructs that live in a self-consistent but adS universe.  Therefore, eternal black holes could be considered "atypical" due to their specific mathematical and physical properties that make them less representative of the black holes in our universe, 

Marolf and Polchinski claim that, under certain high-energy scenarios, black holes formed by collapse can closely resemble eternal black holes when they reach a stationary state. This resemblance might occur within a microcanonical ensemble that specifies a fixed energy range. However, the insights gained from studying AdS eternal black holes might not straightforwardly transfer to the physics of black holes in a dS space, which is a closer representation of our universe. Specifically, eternal black holes may not serve as direct models for understanding black holes in our universe, which is more accurately described by dS space. AdS and dS spaces have fundamentally different geometries, which lead to differing gravitational physics. The behaviors, interactions, and even the formation of black holes can significantly differ between these two settings. Our universe has a positive cosmological constant, characteristic of dS space. This plays a significant role in the large-scale dynamics of the universe, including the behavior of black holes. AdS space, on the other hand, has a negative cosmological constant. Moreover, while AdS/CFT correspondence is a powerful tool within AdS space, no analogous precise duality is known in dS space, making the transition of insights from AdS to dS non-trivial. Given these fundamental differences, insights derived from eternal black holes in AdS space may not hold in dS, and the lack of precise dS/CFT correspondence makes such translations even more challenging.

A Kerr/CFT correspondence was developed to provide a holographic description of extremal rotating (Kerr) black holes by relating the microstates of a black hole to a two-dimensional CFT \cite{Haro}. But extremal black holes have a zero Hawking temperature, so they do not emit Hawking radiation, unlike their non-extremal counterparts. Extremal black holes, with their zero Hawking temperature, do not pose the same informational challenges; the firewall paradox does not arise similarly. In other words, the information and firewall paradoxes primarily concern non-extremal black holes that emit Hawking radiation, leading to the potential loss of information. The Kerr/CFT correspondence focused on extremal (or near-extremal) black holes does not directly address and resolve the firewall and information paradoxes associated with Hawking radiation. Its primary aim is to provide a holographic description to understand the microstate counting and the entropy of such black holes, aiding in the broader understanding of quantum gravitational phenomena in these extreme conditions.

\subsection{The ER = EPR conjecture}
 
In the AdS/CFT correspondence context, a two-sided black hole is an eternal black hole in a maximally extended spacetime, with two exterior regions connected by a wormhole. On the other hand, a one-sided black hole is more like an astrophysical black hole formed from collapse and doesn’t have this sort of maximally extended spacetime. A one-sided black hole initially lacks Hawking radiation but emits Hawking quanta over time. These quanta are entangled with the black hole, which, according to ER $=$ EPR, could imply a kind of wormhole connection. The entanglement entropy increases until it reaches a maximum at the Page time, after which it decreases as more quanta are emitted, signifying a change in the entanglement structure. After the Page time, the Hawking radiation begins to carry away information about the interior of the black hole, as the emitted radiation becomes less entangled with the black hole and more entangled with the earlier emitted radiation. The change in entanglement entropy and the shrinking of the black hole have implications for the smoothness of the horizon. 

In 2013, Maldacena and Susskind suggested that the two-sided black hole is initially similar to the setup where there are two black holes, Alice's and Bob's, connected by a non-traversable ER bridge due to entanglement. However, as Hawking radiation comes into play, Alice's black hole is replaced by this radiation. Their description suggests that the entanglement between the remaining black hole (Bob's) and the Hawking radiation could be depicted as a modified ER bridge with many exits corresponding to the many particles in the Hawking radiation. The transformation of a two-sided black hole into a one-sided black hole plus Hawking radiation, as described, is a way to illustrate the complex and evolving entanglement relationships as Hawking radiation is emitted and interacts with the black hole system \cite{Maldacena}. 

In the Penrose diagram of an eternal black hole, the two sides represent two asymptotically flat or AdS regions of spacetime connected by an ER bridge. Maldacena and Susskind use this diagram to create an analogy: one side (right side) represents Bob’s black hole while the other (left side) metaphorically represents Hawking radiation. This analogy helps to provide a geometric picture of the entanglement between the black hole and its Hawking radiation. 
Maldacena and Susskind argue that this geometric representation of entanglement does not necessitate a firewall at the horizon. The entanglement between Bob’s black hole and the Hawking radiation (interpreted geometrically as a wormhole) maintains the event horizon's smoothness, adhering to general relativity predictions while accounting for the quantum entanglement.
The firewall argument arises primarily in one-sided black holes, like those described by the Schwarzschild solution or other astrophysical black holes, especially when considering the implications of Hawking radiation and the information paradox over long periods. It challenges the classical notion of the no-drama scenario at the event horizon, as predicted by general relativity. However, if the ER $=$ EPR conjecture holds, it implies that the entangled particles created by Hawking radiation are geometrically connected to the one-sided black hole's interior via a wormhole. This setup might offer a way to resolve the monogamy of entanglement problem without invoking a firewall, potentially allowing for the preservation of the equivalence principle \cite{Maldacena}.

\subsection{A firewall is unnecessary}

If there were a firewall, it would be a highly energetic, disruptive boundary at the event horizon of a black hole, which contradicts the predictions of general relativity. General relativity suggests that the event horizon should be smooth and unremarkable from a local perspective. The entanglement between the black hole and the Hawking radiation could lead to a buildup of entanglement entropy, which was feared to result in a firewall. The ER $=$ EPR conjecture potentially resolves this conflict. By interpreting the entanglement between the black hole and the Hawking radiation as a wormhole, Maldacena and Susskind provide a geometric, spacetime-based framework to understand the quantum entanglement without disrupting the smoothness of the event horizon. The wormhole acts as a conduit for entanglement, potentially allowing for information reconciliation without needing a firewall. The geometry of spacetime via wormholes could provide a way to preserve information about particles falling into the black hole, aligning with quantum mechanical principles while maintaining a smooth horizon as per general relativity. 

Maldacena and Susskind mention two possible interpretations of the AMPS argument \cite{Maldacena}: 

1. In the first scenario, Alice can create a high-energy particle (in mode $A$) when she distills information from radiation ($R_B$) far from the black hole. If this is the case, this action alone doesn't disrupt the smoothness of the black hole's horizon—no firewall is formed. In other words, a firewall is unnecessary since the high-energy particle was not present before Alice's action. 

2. In the second scenario, the high-energy particle is present regardless of Alice's actions, implying there is a firewall.

In Maldacena and Susskind’s view, the excitation or the high-energy particle Alice encounters at the horizon results from her interaction with qubit $R_B$ before she falls into the black hole, mediated through a wormhole. The ER $=$ EPR conjecture proposes a connection between quantum entanglement and spacetime geometry, potentially providing a new way to understand the black hole interior. In this picture, the early radiation from the black hole (entangled with the black hole) could provide a complementary description of the black hole's interior. In the case of a black hole, the early radiation emitted by the black hole is entangled with the interior of the black hole. According to ER $=$ EPR, a wormhole might mediate this entanglement. When Alice interacts with a qubit from the early radiation (qubit $R_B$), this interaction might affect the state at the horizon, potentially creating the excitation (or firewall) she encounters as she falls into the black hole. This happens through the wormhole connecting qubit $R_B$ and the black hole's interior. 

ER $=$ EPR posits a potentially smooth interior linked to the exterior by entanglement-mediated wormholes (interpretation 1), while AMPS suggests a disruptive firewall at the horizon (interpretation 2). ER $=$ EPR suggests that interaction with early radiation could change the state at the horizon due to entanglement. In contrast, AMPS argue that the entanglement between early and late radiation creates a contradiction leading to the firewall. 

While ER $=$ EPR may provide a way to reconcile the seemingly smooth nature of the event horizon with quantum entanglement, it's a conjecture that has not been proven. If correct, ER $=$ EPR might eliminate the need for firewalls, providing a smooth spacetime structure inside black holes. However, the exact nature of the black hole interior, the reality of firewalls, and the validity of the ER $=$ EPR conjecture are all open questions requiring further investigation.

The difference between the AMPS claim and Maldacena and Susskind's claim mainly lies in their interpretation of the entanglement scenario and its consequences for the structure of the black hole's event horizon and interior. While AMPS suggest a firewall at the horizon due to a violation of entanglement monogamy, Maldacena and Susskind propose a more harmonious resolution through the ER $=$ EPR conjecture, indicating that the black hole's interior and the early radiation could be two complementary descriptions of the same reality, mediated by quantum entanglement and possibly connected by wormholes.
AMPS argue against the first scenario, stating that Alice couldn’t have created the particle due to the distance of the distillation process from the black hole and the particle’s observed motion. Contrary to AMPS, Maldacena and Susskind argue that Alice created the high-energy particle when she distilled $R_B$. They suggest that a corresponding quantum in mode A is created if Alice distills one such mode, but the horizon remains smooth if no other disturbance occurs, i.e., if Alice does not distill any $R_B$.  

The big question is how Alice could create a particle in mode A from a distance. Maldacena and Susskind propose that the particle reaches mode A not outside the black hole but via an ER bridge. In scenarios where Hawking radiation is replaced by a second entangled black hole, fitting with the ER $=$ EPR conjecture, they feel more confident in this explanation.  
Even though Maldacena and Susskind believe the firewall proposal is incorrect, the challenging concept pushed them to thoroughly explore the nexus between quantum information science and quantum gravity.

While such ideas' mathematical elegance and potential explanatory power are enticing, they remain speculative frameworks within which physicists can explore and resolve the AMPS paradox without empirical verification. But this speculative nature is a known aspect of quantum gravity, especially when delving into realms that are currently beyond the reach of experimental testing and observation. 

Susskind mentions that if there are firewalls at the horizons of black holes, they could straightforwardly resolve the paradox without necessitating a rethinking of quantum mechanics. However, the mechanism through which these firewalls would form is yet to be identified or understood. This proposition preserves the standard interpretations of quantum mechanics but at the cost of introducing a radical new phenomenon at the event horizons of black holes. On the other hand, the ER $=$ EPR conjecture leads to a more profound and potentially revolutionary understanding of quantum mechanics, tying together quantum entanglement and geometric connectivity \cite{Susskind4}. Yet, this idea also hinges on wormholes, theoretical constructs that haven't been observed. 

As discussed in section \ref{3}, Harlow and Hayden approach the firewall problem using computational considerations. The Harlow-Hayden conjecture provides a computational complexity-based resolution to the AMPS paradox. They propose that recovering information from a black hole by manipulating the Hawking radiation is computationally infeasible within the time frame before the black hole evaporates. This approach ties more closely to information theory and computational complexity, providing a framework that may feel more grounded to some.

Maldacena and Susskind, on the other hand, delve into the realm of wormholes and ER bridges to propose a geometric interpretation of entanglement, aiming to resolve the paradox by connecting quantum entanglement with spacetime geometry. While the Harlow-Hayden approach might seem more accessible or realistic due to its grounding in computational theory, it's still within a theoretical realm yet to be empirically validated. Furthermore, the acceptance of one approach over the other could be influenced by advancements in theoretical understanding or experimental techniques that could provide evidence supporting one conjecture.

\subsection{Causality and chronology protection as a barrier} 

According to Harlow and Hayden, a key obstruction to the AMPS experiment is the high computational complexity required to distill purification from Hawking radiation, which is likely to take an exponential amount of time relative to the black hole's entropy, far exceeding the black hole's evaporation time.

Susskind suggests that there could be a principle other than computational complexity that prevents the experiment from being conducted, specifically, causality. Hawking's chronology protection conjecture argues against the physical plausibility of time loops (or closed timelike curves) in nature, which would allow for causality violations akin to time travel. The narrative wherein Alice extracts $R_B$ (a representation of future quantum state $A$) from early radiation and reintroduces it to the black hole to meet $A$ is analogized to time travel. It's as if $A$ has traveled back in time as $R_B$ met its future self, creating a closed causal loop. In the AMPS scenario, introducing a firewall—representing a dramatic deviation from the expected smooth spacetime near the horizon—breaks this loop and enforces chronology protection, similar to the idea behind Hawking's conjecture. The firewall ensures that the entanglement arrangement leading to a causality violation is interrupted, thus maintaining the causal structure of spacetime. 

Harlow and Hayden proposed that performing this distillation task with a quantum computer would require time exponential in $N$, where $N$ is a parameter related to the system size or the number of qubits involved. This exponential time complexity presents a significant computational hurdle as it means that the time required to complete the distillation grows exponentially with the size of the system. If the conjecture holds, Alice could not complete the distillation of $R_B$ until after the black hole has evaporated. This timeline implies that by the time she could verify the entanglement between $R_B$ and $B$, the black hole itself, and thus state A, would no longer exist. 

This scenario effectively acts as a form of chronology protection, ensuring that the causality-violating scenario envisaged in the AMPS paradox cannot occur. The chronology protection mechanism here is not a physical barrier like a firewall but rather a computational limitation that prevents Alice from extracting and utilizing the information in $R_B$ in a way that would lead to a causality contradiction.

In the case of black holes within AdS space, the evaporation process is arrested, giving Alice unlimited time to perform her distillation task. Yet, the AMPS paradox still hints at a necessity for a barrier (a computational barrier) to prevent causality contradictions, rendering the barrier as a mechanism for chronology protection even in such non-evaporating scenarios: The mechanism of chronology protection here is not a physical barrier like a firewall, but rather a computational limitation that prevents Alice from extracting and utilizing the information in $R_B$ in a way that would lead to a causality contradiction. The alignment of the evaporation timeline of the black hole with the exponential time complexity of the distillation task serves as a synchronicity that preserves the causal order of events and avoids the potential paradoxes that would arise if Alice could complete the distillation before the black hole evaporates. 

Susskind argues that the Harlow-Hayden conjecture introduces a computational barrier to the hypothetical sequence of events posed in the AMPS paradox, serving as chronology protection through the inherent limitations of quantum computational processes \cite{Susskind},\cite{Susskind3}, \cite{Susskind5}.

\section{A zoo of proposals} \label{5} 

\subsection{A shift in perspective}

Einstein spent a significant portion of his later life attempting to formulate a Unified Field Theory where he aimed to unify gravity with electromagnetism within the framework of General Relativity. His approach was largely deterministic and geometrical, attempting to encompass quantum mechanics within a general relativistic framework \cite{Sauer}. 

Post advancements in understanding quantum mechanics, modern efforts have largely aimed at incorporating gravity into the quantum mechanical framework. The prevailing belief among many physicists is that a quantum theory of gravity is the necessary next step. This is a stark shift from Einstein's geometrical, classical approach towards a probabilistic, quantum approach. One promising pathway is the AdS/CFT correspondence, a part of the larger holographic principles and string theory framework. 

AdS/CFT suggests an emergent nature of gravity and spacetime geometry from a lower-dimensional quantum theory, contrasting with Einstein's view of gravity as a fundamental force described geometrically. This reverses Einstein's approach, focusing on deriving gravitational phenomena from quantum mechanics rather than vice versa. Modern approaches open the door to describing physical phenomena in terms of quantum information and computation by rooting the foundations of physics in quantum mechanics. This is not easily accomplished if starting from a classical general relativistic foundation. Unitary transformation is aligned with the quantum mechanical view of the universe’s evolution. In quantum mechanics, the evolution of systems is described by unitary transformations, which preserve the total probability (or information) within the system, further tying into the information theoretical description of the universe.

The change in approach over the decades reflects the accumulation of experimental evidence supporting quantum mechanics and the difficulties encountered in attempting to unify physics starting from a classical, geometrical foundation as Einstein had attempted. This shift towards quantum foundations and the exploration of emergent gravity through frameworks like AdS/CFT represents a profound change in our conceptual understanding of the fundamental forces of nature and how they might be unified. 
In the AdS/CFT correspondence context, the boundary CFT is a quantum field theory that obeys the unitary evolution principles. So, if the duality holds, then unitarity should also be preserved in the bulk gravitational theory, and thus, there should be no information loss in black holes. While the advent of the AdS/CFT
correspondence brought a new perspective, and despite its allure, it confronts challenges in its direct application to our universe’s dS space-time.

\subsection{Fuzzballs, Remnants, and the final burst}

In 2017, Unruh and Wald addressed a long-standing debate in physics about whether information can truly be lost inside a black hole \cite{Unruh}. Over the years, many alternatives to this idea of information loss have been proposed \cite{Unruh}:

1. \emph{Fuzzballs}: the fuzzball concept suggests that during the gravitational collapse of a star, instead of forming a traditional black hole with an event horizon, some other structure, termed a fuzzball, might form \cite{Mathur}. 
The Fuzzball proposal implies that since there's no event horizon, there's no associated information loss problem. Therefore, the challenge posed by the black hole information paradox would not arise.

Unruh and Wald challenge the fuzzball conjecture based on inconsistencies. They see the fuzzball idea as a significant departure from classical understandings of physics. In classical general relativity, they highlight that sufficiently massive black holes can form even at very low energy densities and curvatures. For a fuzzball to form instead, there would need to be a severe breakdown of established physics in areas where these models should work well. They argue that if a massive shell collapses inwards, seeing how it could stop its inward momentum just in time to form a fuzzball rather than a traditional black hole is challenging. This would seem to go against principles like momentum conservation; for a fuzzball to form, extreme and sudden stresses would need to appear, which would cause significant alterations to the effective spacetime metric. These sudden stresses would not align with the energy densities, creating a scenario that is hard to reconcile with current physical understanding. Classical black holes and their horizons are global structures. If the collapsing matter were to behave differently to form a fuzzball and avoid forming a horizon suddenly, it would imply that the matter has some form of non-local knowledge – which again challenges conventional physics. 
Unruh and Wald find it difficult to reconcile such ideas with known physics \cite{Unruh}. 

2. \emph{Remnants}: Hawking's calculation suggested that black holes evaporate completely. However, this calculation is based on assumptions that may not hold as the black hole approaches the Planck scale. The question remains whether a black hole can fully evaporate or if it would leave a \emph{remnant} behind. Some calculations hint at such remnants, but a definitive answer remains elusive due to the lack of a complete quantum gravity theory. Since there is no apparent rule preventing a black hole from fully evaporating, it should do so, behaving like any unstable quantum system. However, the advocates of the \emph{remnants} conjecture argue against complete evaporation, drawing a parallel with the hydrogen atom. The atom does not collapse because of the uncertainty principle. Similarly, a generalized uncertainty principle may prevent black holes from evaporating completely. This principle extends the standard uncertainty principle, incorporating quantum gravitational effects. In this model, the black hole ceases to radiate once it approaches the Planck mass, leaving a stable remnant behind. It is suggested that such remnants might be a candidate for dark matter \cite{Adler}.

3. Finally, according to another suggestion, the \emph{final burst} idea, as a black hole evaporates and reduces to the Planck scale, all the information stored within the black hole is suddenly and explosively released, leading to a pure final state. Although it seems counterintuitive for a Planck-sized object to release massive amounts of information, Unruh and Wald suggest looking into models where such behavior could be plausible. 

Unruh and Wald raise a key question concerning the remnants conjecture: Can these Planck scale remnants interact with the external universe? If remnants do not interact, they contend that the information within the remnants would be trapped and inaccessible. This would imply that, for all practical purposes, the system's final state would still be mixed, rendering the remnant somewhat moot in the context of the information paradox.

At first glance, the idea of the final burst might seem implausible because it implies that an enormous amount of information is released from an object that has reduced to the Planck size, which is extremely small. Normally, one would expect that releasing a large amount of information would require the release of a correspondingly large amount of energy. However, just like the remnant proposal, in the final burst conjecture, the final burst is not considered radical because it suggests a deviation from our current understanding of black hole physics only when the black hole reaches the Planck scale. Our semiclassical understanding of gravity is expected to break down at this scale, so postulating new behaviors here is not considered extreme. 
Unruh and Wald conclude that the final burst proposal remains a viable alternative to explain the black hole information paradox \cite{Unruh}.

Unruh and Wald point out that the evolution of a black hole from a pure state to a mixed state (i.e., information loss) violates the unitarity of quantum mechanics. 
However, they also argue that the evolution from a pure state to a mixed state, as suggested by semiclassical analyses of black hole evaporation, is not a violation of quantum theory but can be a prediction in certain contexts. 
While highlighting the tension between information loss and certain principles of quantum mechanics, they see scenarios where quantum theory allows for such evolution (from pure to mixed states) without being violated. Unruh and Wald advocate for a nuanced view, emphasizing the robustness of quantum mechanics, even if it means accepting challenging results. 

\subsection{The Page time paradox}

In 2018, David Wallace introduced the \emph{Page Time Paradox}, a variation of the information paradox \cite{Wallace}. The Page time paradox arises from the contrast between the predictions of quantum field theory (QFT) and the predictions of black hole statistical mechanics in the scenario of a black hole radiating away its energy, as described by Hawking radiation. 

Wallace considers a typical thermodynamic system, i.e., a black hole, with a defined Hilbert space. The black hole is initially in a pure state at a certain energy level, $E_0$, and it cools down by emitting thermal radiation. The radiation is without memory of the specific state of the black hole. While the radiation is thermal, each emitted quantum will be in a mixed state. To preserve the overall purity of the black hole-plus-radiation state, each quantum of radiation must be entangled with the black hole. According to unitarity, the total von Neumann entropy of the emitted radiation as the black hole cools down should be balanced by an equal von Neumann entropy of the black hole to keep the total state of black hole-plus-radiation pure. This suggests that the black hole becomes mixed to balance the entropy as radiation with mixed states is emitted. As the black hole cools down and transitions through different energy levels, the microcanonical entropy at each level provides a bound for the von Neumann entropy the black hole can have at that level. This is due to the definition of the microcanonical ensemble, which assumes a fixed energy level and an equal probability of being in any microstate within that energy level.

Over time, as more radiation is emitted, a conflict arises. The unitarity principle suggests the black hole’s von Neumann entropy should increase to balance the entropy of the emitted radiation. However, microcanonical entropy decreases as the black hole energy decreases. Microcanonical entropy bound at lower energy levels limits how much von Neumann entropy the black hole can have. The Page Time is the point at which these two effects balance each other. Importantly, the Page time signals when half of the black hole's entropy has been radiated away. After this time, for unitarity to be preserved, the radiation must carry away more information, i.e., it can't be purely thermal. This transition from purely thermal to more informative radiation is central.

After this point, the emitted radiation can no longer be purely thermal, as that would require the black hole’s von Neumann entropy to exceed the microcanonical entropy bound. To resolve this, the later emitted radiation must be entangled with the earlier emitted radiation, reducing the overall von Neumann entropy and adhering to the microcanonical entropy bound. However, it is crucial to note that the QFT calculations giving rise to Hawking radiation as purely thermal are done in a semi-classical regime, which combines classical general relativity with QFT rather than a full quantum gravity calculation. This semi-classical calculation shows the radiation as exactly thermal, with no such entanglement, thus contradicting the predictions from black hole statistical mechanics. Wallace called this discrepancy the \emph{Page-Time Paradox} \cite{Wallace}. 

In an interview, Wallace said: "The Page-time paradox seems to point to a breakdown of low-energy physics in a place where it has no business breaking down, because the energies are still low" \cite{Musser}. Wallace's statement underscores the unexpected nature of the paradox. When we think of breakdowns in physics, we imagine extreme conditions, e.g., very high energies. However, the Page-time paradox arises even when the energies involved are still relatively low. It is puzzling because, under such low-energy conditions, QFT, a standard theory, should be perfectly valid. The appearance of this paradox in such conditions hints at a deeper problem or a missing piece in our understanding of black hole physics.

\subsection{Are we in the midst of a crisis?}

There is a certain historical cyclicality in the foundations of physics, where different theoretical frameworks vie for dominance and a unified understanding. The historical context of the $19^{\textbf{th}}$ century among the mechanistic, energetic, and electromagnetic worldviews mirrors the current struggle between general relativity and quantum mechanics frameworks in some ways. The competing frameworks in the $19^{\textbf{th}}$ century tried to encapsulate all known physical phenomena within a coherent theoretical structure. Each had its base entity, ether, energy, or mechanical matter, from which other phenomena were to be derived. In the late $19^{\textbf{th}}$ and early $20^{\textbf{th}}$ centuries, paradoxes and anomalies were resolved and clarified through the advent of the theories of relativity and quantum mechanics. Similarly, general relativity and quantum mechanics, each extraordinarily successful in its realm, face challenges regarding new paradoxes and anomalies. 

Thomas Kuhn discerns a cyclical pattern in the history of science. Kuhn’s cyclical picture of scientific revolutions consists of the following key elements \cite{Bird}:

1. \emph{Normal Science}: This phase is characterized by the widespread acceptance and practice of a particular paradigm and scientific ideas and methods. Scientists work on puzzle-solving within this paradigm, further refining and expanding it.

Certain experimental results and phenomena emerge that don't fit well within the current paradigm over time. Initially, these are treated as anomalies and exceptions.
Kuhn distinguishes between a \emph{severe anomaly} and an \emph{ordinary anomaly}. Kuhn stipulates that a \emph{severe anomaly} is recognized when "all members of a scientific group will reach the same decision" \cite{Kuhn}. Based on this criterion, the contentious debate surrounding the black hole information paradox and its manifestation in the firewall paradox can be aptly categorized as a severe anomaly.

2. \emph{Crisis}: As more anomalies accumulate and the existing paradigm struggles to account for them, there is growing doubt and a sense of crisis in the scientific community. 

Just as political conflicts can be resolved within a constitution, normal science resolves anomalies within a paradigm. However, when the system becomes a problem, a revolution becomes necessary. 

3. \emph{Revolution}: The crisis eventually leads to searching for a new paradigm to explain the anomalies better. Kuhn calls this shift from one paradigm to another a scientific revolution. It is a dramatic, paradigm-shifting change in the field.
The shift between paradigms is not always smooth. Just as political revolutions require force and propaganda, scientific revolutions involve external factors and individual biases.

4. \emph{New Paradigm}: Accepting a new paradigm ushers in another period of normal science, and the cycle begins anew: normal science, crisis, revolution, new paradigm, normal science.
According to Kuhn, there isn't always a clear translation between the language and concepts of different paradigms. This leads to the idea that paradigms are incommensurable and fundamentally incomparable. A paradigm change is likened to a gestalt shift. Scientists operating within different paradigms might see the world in fundamentally different ways. 

The meta-irony in Kuhn's ideas is that if scientific paradigms are indeed subject to cycles of acceptance, crisis, and revolution, then Kuhn's paradigm about the nature of scientific revolutions might be subject to the same cycle. Applying Kuhn's model to his theory underscores the potential universality of his insights. This suggests that no theory, including Kuhn's, is immune from being revised, replaced, or overthrown as new insights and anomalies emerge. It's a humbling reminder of the evolving nature of knowledge and understanding.

\section*{Acknowledgement}

\noindent This work is supported by ERC advanced grant number 834735.


\begin{thebibliography}{38}

\bibitem[Ad-Ch-Sa]{Adler} Adler, R. J., Chen, P., and Santiago, D. I., (2022). "The Generalized Uncertainty Principle and Black
Hole Remnants." \emph{General Relativity and Gravitation} 33, pp. 2101-2108.

\bibitem[AEHPS]{Akers} Akers, C., Engelhardt, N., Harlow, D., Penington, G., and Vardhan, S. (2022). "The black hole interior from non-isometric codes and complexity." \emph{arXiv}:2207.06536v2 [hep-th].

\bibitem[AMPS]{AMPS} Almheiri, A., Marolf, D., Polchinski, J., and Sully, J., (2013). "Black Holes: Complementarity or Firewalls?" \emph{Journal of High Energy Physics} 2013, pp. 1-19.

\bibitem[AMPSS]{AMPSS} Almheiri, A., Marolf, D., Polchinski, J., Stanford, D., and Sully, J., (2013). "An Apologia for Firewalls." \emph{Journal of High Energy Physics} 2013, pp. 1-31.

\bibitem[Bir]{Bird} Bird, A. (2000). \emph{Thomas Kuhn}. Chesham: Acumen Publishing Limited.

\bibitem[HMB]{Haro} De Haro, S., Mayerson, D. R., Butterfield, J. N. (2016). "Conceptual Aspects of Gauge/Gravity Duality." \emph{Foundations of Physics} 46, pp. 1381–1425.

\bibitem[Har-Hay]{Harlow} Harlow, D., and Hayden, P. (2013). “Quantum computation vs. firewalls.” \emph{Journal of High Energy Physics} 2013, pp. 1-56.

\bibitem[Har]{Harlow2} Harlow, D. (2016). “Jerusalem lectures on black holes and quantum information.” \emph{Reviews of Modern Physics} 88, pp. 015002-1- 015002-58.

\bibitem[Hart]{Hartman} Hartman, T., (2015). "Lectures on Quantum Gravity and Black Holes." \emph{CornellPhysics7661}.

\bibitem[Hay-Pres]{Hayden} Hayden, P., and Preskill, J. (2007). "Black holes as mirrors: quantum information in
random subsystems". \emph{Journal of High Energy Physics} 2007, pp. 1-20. 

\bibitem[Hawk1]{Hawking} Hawking, S. (2005). "The Information Paradox for Black Holes." \emph{arXiv}:1509.01147v1 [hep-th]. 

\bibitem[Hawk2]{Hawking2} Hawking, S. (2005). "Information Loss in Black Holes." \emph{Physical Review D} 72, pp. 084013-1-084013-4. 

\bibitem[Kap]{Kaplan} Kaplan, J. (2016). "Lectures on AdS/CFT from the Bottom Up." \emph{Johns Hopkins University}.

\bibitem[Kuhn]{Kuhn} Kuhn, T. (1970). "Reflections on my Critics" In Lakatos, I. and Musgrave, A. (eds.). \emph{Criticism and the Growth of Knowledge. Proceedings of the International Colloquium in the Philosophy of Science, London, 1965, volume 4}. Cambridge: Cambridge University Press, pp. 231-278.

\bibitem[LPSTU]{Susskind2} Lowe, D. A., Polchinski, J., Susskind, L., Thorlacius, L. and Uglum, J. (1995). “Black Hole Complementarity versus Locality.” \emph{Physical Review D} 52, 6997.

\bibitem[Mald-Suss]{Maldacena} Maldacena, J. and Susskind, L. (2013). “Coll horizons for entangled black holes,” \emph{Fortschritte der Physik} 61, pp. 781-811. 

\bibitem[Pol-Mar]{Polchinski2} Marolf, D., and Polchinski, J. (2013). "Gauge/Gravity Duality and the Black Hole Interior." \emph{Physical Review Letters} 111, pp. 171301-1-171301-5.

\bibitem[Math]{Mathur} Mathur, S. D. (2005). "The fuzzball proposal for black holes: an elementary review." \emph{Fortschritte der Physik} 53, pp. 793-827. 

\bibitem[Mus]{Musser} Musser, G. (2020)."The Most Famous Paradox in Physics Nears Its End." \emph{QuantaMagazina}. 

\bibitem[Par-Wil]{Wilczek} Parikh, M.K., and Wilczek, F. (1998). "An Action for Black Hole Membranes." \emph{Physical Review D} 58, pp. 064011-1-064011-12.

\bibitem[Pol]{Polchinski} Polchinski, J. (2016). “The Black Hole Information Problem.” \emph{arXiv}:1609.04036v1 [hep-th].

\bibitem[Pr-Th]{Thorne} Price, R. H., and Kip S. T., (1988). "The Membrane Paradigm for Black Holes." \emph{Scientific American} 258, pp. 69-77.

\bibitem[Sau]{Sauer} Sauer, T. "Einstein’s Unified Field Theory Program." In Janssen, M. and Lehner, C. (2014). \emph{The Cambridge Companion to Einstein}. Cambridge: Cambridge University Press.

\bibitem[Sek-Suss]{Sekino}Sekino, Y., and Susskind, L. (2008). “Fast Scramblers.” \emph{Journal of High Energy Physics} 10, pp. 1-14. 

\bibitem[Suss95]{Susskind6} Susskind, L. (1995). "The World as a Hologram." \emph{Journal of Mathematical Physics} 36, pp. 6377-6396.

\bibitem[Suss12]{Susskind} Susskind, L. (2012). “The Transfer of Entanglement: The Case for Firewalls,” \emph{arXiv}:1210.2098v1 [hep-th].

\bibitem[Suss13]{Susskind3} Susskind, L. (2013). “Black Hole Complementarity and the Harlow-Hayden Conjecture,” \emph{arXiv}:1301.4505v2 [hep-th].

\bibitem[Suss16]{Susskind4} Susskind, L. (2016). “ER = EPR, GHZ, and the consistency of quantum measurements.” \emph{Fortschritte der Physik} 64, pp. 72-83.

\bibitem[Suss20]{Susskind5} Susskind, L. (2020). \emph{Three Lectures on Complexity and Black Holes}. Switzerland: Springer. 

\bibitem[STU]{Susskind1} Susskind, L., Thorlacius, L. and Uglum, J. (1993). “The Stretched Horizon and Black Hole Complementarity.” \emph{Physical Review D} 48, pp. 3743-3762.

\bibitem[Pr-Th-Mac]{Thorne2} Thorne, K.S., Price, R.H., and Macdonald, D.A., (1986). \emph{Black Holes: The Membrane Paradigm.} New Haven: Yale University Press.

\bibitem[Un-Wal]{Unruh} Unruh, W. G., and Wald, R. M. (2017) “Information Loss.” \emph{Reports on Progress in Physics} 80, 092002. 

\bibitem[Wall18]{Wallace} Wallace, D. (2018). "Why black hole information loss is paradoxical." \emph{arXiv}:1710.03783v2 [gr-qc].

\bibitem[Wall-mem.18]{Wallace1} Wallace, D. (2018). "The case for black hole thermodynamics part I: Phenomenological thermodynamics." \emph{Studies in History and Philosophy of Modern Physics} 64, pp. 52-67.

\bibitem[Wall-mem.19]{Wallace2} Wallace, D. (2019). "The case for black hole thermodynamics part II: Statistical mechanics." \emph{Studies in History and Philosophy of Modern Physics} 66, pp. 103-117.

\bibitem[Win]{Weinstein} Weinstein, G. (2013). "Reframing the Event Horizon: The Harlow-Hayden Computational Approach to the Firewall Paradox." \emph{arXiv}:2309.09382.


\end{thebibliography}
\end{document}